# Information theoretically secure, enhanced Johnson noise based key distribution over the smart grid with switched filters

Elias Gonzalez, Laszlo B. Kish[*], Robert S. Balog, Prasad Enjeti

Department of Electrical and Computer Engineering, Texas A&M University, College Station, TX 77843-3128, USA
*E-mail: Laszlo.Kish@ece.tamu.edu

## Abstract

We introduce a protocol with a reconfigurable filter system to create non-overlapping single loops in the smart power grid for the realization of the Kirchhoff-Law-Johnson-(like)-Noise secure key distribution system. The protocol is valid for one-dimensional radial networks (chain-like power line) which are typical of the electricity distribution network between the utility and the customer. The speed of the protocol (the number of steps needed) versus grid size is analyzed. When properly generalized, such a system has the potential to achieve unconditionally secure key distribution over the smart power grid of arbitrary geometrical dimensions.

## Introduction

### 1.1 KLJN, the information theoretically secure wire-based key exchange scheme

On February 12, 2013 President Obama issued an executive order to outline policies directing companies and operators of vital infrastructure, such as power grids, to set standards for cybersecurity [1]. This step is one of the indications of an urgent need to protect intelligence, companies, infrastructure, and personal data in an efficient way. In this paper, we propose a solution that provides information theoretic (that is, unconditional) secure key exchange over the smart grid. This method is controlled by filters and protects against man-in-the-middle attacks.

A smart grid [2,3] is an electrical power distribution network that uses information and communications technology to improve the security [4,5], reliability, efficiency, and sustainability of the production and distribution of electricity. A form of a cyber-physical system, the smart grid enables greater efficiency through a higher degree of awareness and control but also introduces new failure modes associated with data being intercepted and compromised.

Private key based secure communications require a shared secret key between Alice and Bob who may communicate over remote distances. In today's secure communications, sharing such a key also utilizes electronic communications because courier and mail services are slow. However the software based key distribution methods offer only limited security levels that are only *computationally-conditional* thus they are *not future-proof*. By having a sufficiently enhanced computing power, the eavesdropper (Eve) can crack the key and all the communications that are using that key. Therefore, unconditional security (indicating that the security holds even for infinite computational power), which is the popular wording of the term "information theoretic security" [6], requires more than a software solution. It needs the utilization of the proper laws of physics.

The oldest scheme that claims information theoretic security by utilizing the laws of physics is quantum key distribution (QKD). While, the security available in QKD schemes have currently been debated/compromised [7-21], the discussions indicate that there is a potential to reach a satisfactory security level in the future though they may require a new approach [7]. However, current QKD devices are prohibitively expensive and have other practical weaknesses, such as they are sensitive to vibrations, bulky, limited in range, and require a special "dark optical fiber" cable with sophisticated infrastructure.

On the other hand, the smart grid offers a unique way of secure key exchange because each household (host) in the grid is electrically connected. To utilize a wire connection for secure key exchange, a different set of the laws of physics (not the laws applied for QKD that work with optical fibers) must be utilized. Recently a classical statistical physical alternative to QKD, the Kirchhoff-Law-Johnson-(like)-Noise (KLJN) key exchange system has been proposed [22-28], which is a wire-based scheme that is free from several weaknesses of QKD. A recent survey is given in [24]. Similarly to QKD, KLJN is also an information theoretically secure key distribution [24]; however it is robust; not sensitive to vibrations; it has unlimited range [25]; it can be integrated on chips [26]; it can use existing wire infrastructure such as power lines [27]; and KLJN based networks can also be constructed [28].

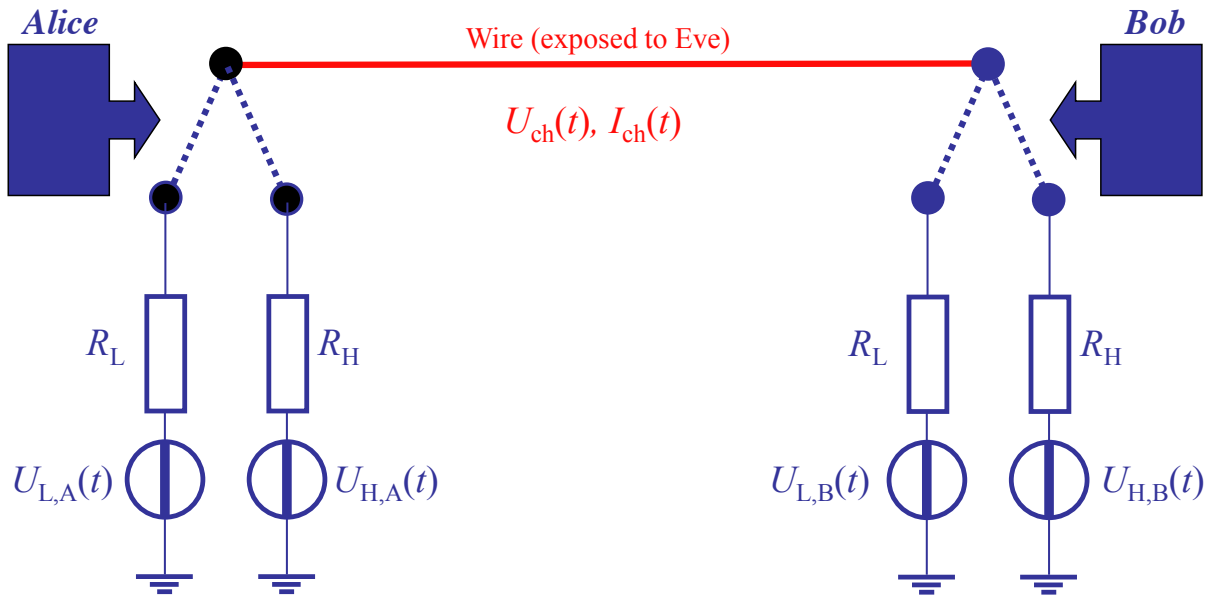

**Figure 1.** The core of the idealized KLJN key scheme [22-25]. This simple system is secure only against passive attacks in the idealized case (mathematical limit). Security enhancements (including filters) to provide protection

against invasive attacks [22-25] and against other types of vulnerabilities are not shown. In practical applications electronic noise generators emulate an enhanced Johnson noise with a publicly agreed high effective temperature.

The KLJN channel is a wire [24]. At the beginning of each clock cycle (note, the 50Hz / 60Hz AC grid provide a universal time synchronization), Alice and Bob, who have identical pairs of resistors $R_L$ and $R_H$ (representing the 0 and 1 bit situations) randomly select and connect one of the resistors, see Figure 1. In practical applications, voltage noise generators enhance the Johnson noise of the resistors so that all resistors in the system have the same, publicly known effective noise-temperature $T_{eff}$ (where $T_{eff} \geq 10^9$ Kelvin). The enhanced Johnson noise voltages $\{U_{L,A}(t)$ or $U_{H,A}(t)$; and $U_{L,B}(t)$ or $U_{H,B}(t)\}$ of the resistor result in a channel noise voltage between the wire and the ground, and a channel noise current $I_{ch}(t)$ in the wire. Low-pass filters are used because the noise-bandwidth, which we also call KLJN-band $B_{kljn}$ (its value depends on the range), must be chosen so narrow that wave, reflection, and propagation/delay effects are negligible, otherwise the security is compromised [22]. Alice and Bob can measure the mean-square amplitudes $\langle U_{ch}^2(t) \rangle$ and/or $\langle I_{ch}^2(t) \rangle$ within the KLJN-band in the line. From any of these values, the loop resistance can be calculated [22] by using the Johnson noise formula with the noise-bandwidth $T_{eff}$:

$$\langle U_{ch}^2(t) \rangle = 4kT_{eff}R_{loop}B_{kljn} \qquad \langle I_{ch}^2(t) \rangle = \frac{4kT_{eff}B_{kljn}}{R_{loop}} \qquad (1)$$

Alice and Bob know their own choice resistor thus, from the loop resistance, they can deduce the resistance value and the actual bit status at the other end of the wire. In the ideal situation, the cases $R_L | R_H$ and $R_H | R_L$ represent a secure bit exchange event because they cannot be distinguished by the measured mean-square values. Eve can do the very same measurements but she has no knowledge about any of the resistance choices thus she is unable to extract the key bits from the measured loop resistance.

The KLJN protocol can also be applied to several other wire networks such as electronic equipment that do not desire to be reverse engineered. However, in this study we focus on applying KLJN protocol on the smart grid.

### 1.2 Utilizing the smart power grid for information theoretic secure key exchange

The disadvantage of the KLJN key exchange protocol is that it requires a wire connection. Investors are hesitant to cover the cost of new infrastructure for solely the purpose of security. On the other hand, virtually every building in the civilized world is connected to the electrical power grid. This fact is very motivating to explore the possibility of using the power grid as the infrastructure for the KLJN protocol. However, only the single loop shown in Figure 1 is unconditionally secure. When Alice and Bob are two remote hosts in the smart grid, they should indeed experience a single loop connection as in Figure 1. Thus, for smart grid applications, proper filters must be installed and controlled for the KLJN frequency band where the key exchange operates. Though simple examples have been outlined to prove that a KLJN key exchange between two remote points in a radial power grid with filters [27,28] can be achieved, neither details about the structure of the filter units nor network protocols to connect every host on the grid with all other hosts have been shown. The method described in [28] is of high speed because, if the units do simultaneous key exchange they have a joint network key. Thus the units must trust each other. In the present system the units have independent keys.

The present paper aims to make the first steps in this direction by presenting a working scheme with scaling analysis of the speed of key exchange versus network size. We limit our network to a one-dimensional linear chain network to utilize the smart power grid for KLJN secure key exchange. We show and analyze a protocol to efficiently supply every host with proper secure keys to separately communicate with all the other hosts.

## Discussions and Results

Because the pattern of connections between KLJN units must be varied to provide the exchange of a separate secure key for each possible pair of host, the network of filters and their connections must be varied accordingly. The power line filter technology is already available [29,30] and we will show that the required results can be achieved by switching on/off proper filtering units, in a structured way in the smart grid. We will need filters to pass or reject the KLJN frequency band $B_{kljn}$ and/or the power frequency (50 or 60 Hz). When both $B_{kljn}$ and $f_p$ are passed, it is a short; and when both of them are rejected, it is a break. We will call these filters "switched filters".

### 2.1. Switched Filters

We call the functional units connected to the smart power grid *hosts*. A host is able to execute a KLJN key exchange toward left and right in a simultaneous way. That means each host has two independent KLJN units. The filter system must satisfy the following requirements:

i) Hosts that currently do not execute KLJN key exchange should not interfere with those processes even if the KLJN signals pass through their connections.

ii) Moreover, each host should be able to extract electrical power from the electric power system without disturbing the KLJN key exchanges.

We define the size of a network as being of size $N$ when that network has $N+1$ hosts. An example of a network of size $N = 7$ is illustrated in Figure 2.

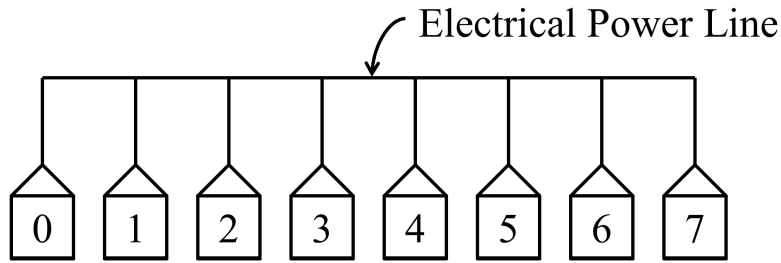

**Figure 2.** Example of a one-dimensional grid: a chain network. This example has a network of size $N = 7$.

Intermediate hosts in the network can be in two different states according to the need:

α) State 1 is defined when KLJN bandwidth $B_{kljn}$ is not allowed into the host.

β) State 2 is defined when KLJN bandwidth $B_{kljn}$ is allowed into the host.

Hosts at the two ends can be in similar situations except that they can communicate in only a single directions, thus they are special, limited cases of the intermediate hosts to which we are focusing our considerations when discussing filters.

*Filter boxes* at each host will distribute the KLJN signals and the power, and they are responsible for connecting the proper parties for the KLJN key exchange and to supply the hosts with power, see Figure 3. The filters boxes can be controlled either by a central server and/or an automatic algorithm. In the following section we discuss the protocol of this control. Each filter box has three switched filters and a corresponding output wire, see Figure 3:

a) The Left KLJN Filter for the KLJN key exchange toward left,

b) The Right KLJN Filter for the KLJN key exchange toward right.

c) The Power Filter to supply power to host.

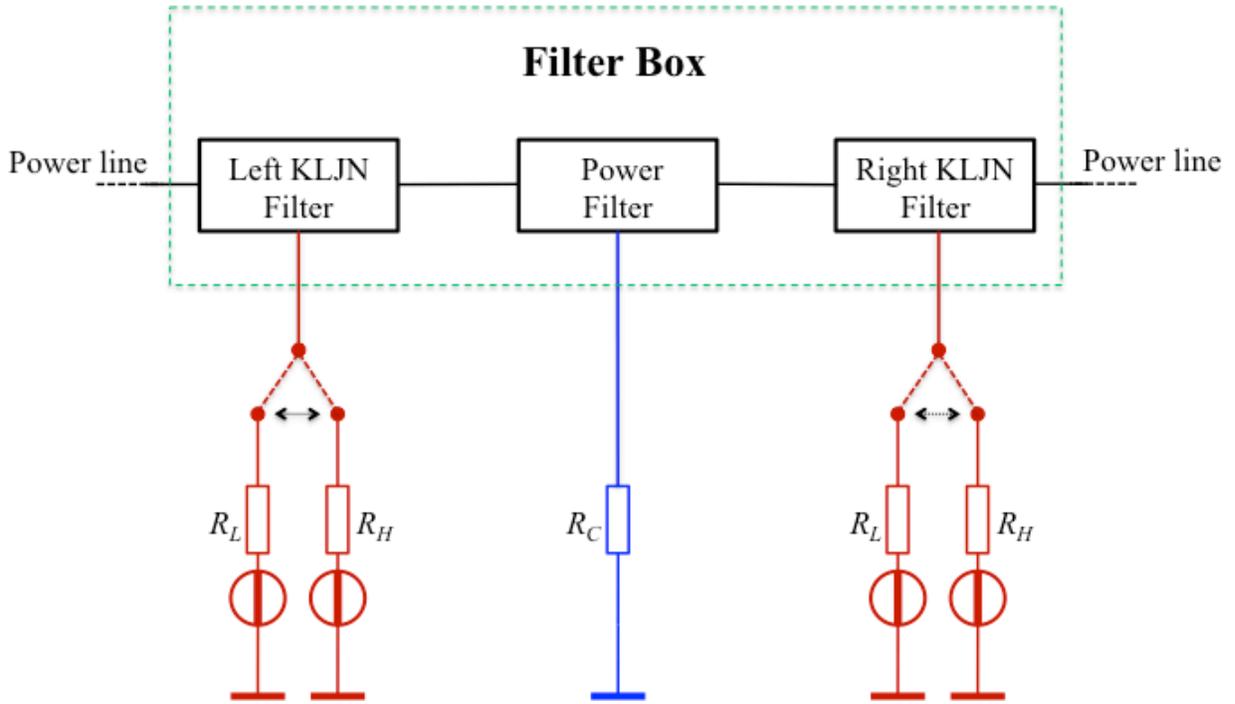

**Figure 3.** Building blocks in a filter box.

The properly controlled filter boxes will provide non-overlapping KLJN loops between the hosts, see below. KLJN loops need to be non-overlapping loops because the KLJN protocol is fundamentally peer-to-peer. If overlapping loops were allowed then there is a possibility that Eve might be in between and will require the trust of the intermediate hosts. A problem with peer-to-peer networks is that they require direct connections. QKD also require direct connections. The reason for having two KLJN units per host is to decrease the time needed to connect every host by having simultaneous loops toward left and right, without overlapping. Figure 4 shows an example for $N = 7$. The solid black line means that both KLJN bandwidth and power frequency is passing through (ordinary wire: the original line). The (red) dotted lines carry $B_{kljn}$ ($f_p$ is rejected). The (blue) dashed lines indicate the opposite situation: only the power frequency is passing and the KLJN bandwidth is rejected.

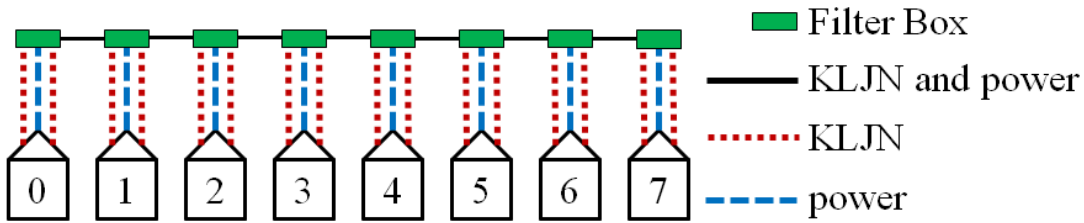

**Figure 4.** Example for network of size $N = 7$. Each host is connected to a filter box and the filters boxes are connected to the power grid. Note how each host has three wire connections to its filter box.

In Figure 13 there is one key exchange between the first host (host 0) and the last host (host 7) in the network, and all host in between (host 1 through host 6) are not allowed to access the KLJN band. In this state the filter boxes of host 1 through 6 must separate their respective host from the KLJN band, and at the same time supply them by power. We call this working mode of the filter boxes of non-active hosts *State 1*. The wiring and frequency transfer of the Filter Box in State 1 are shown in Figure 5 and Tables 1,2.

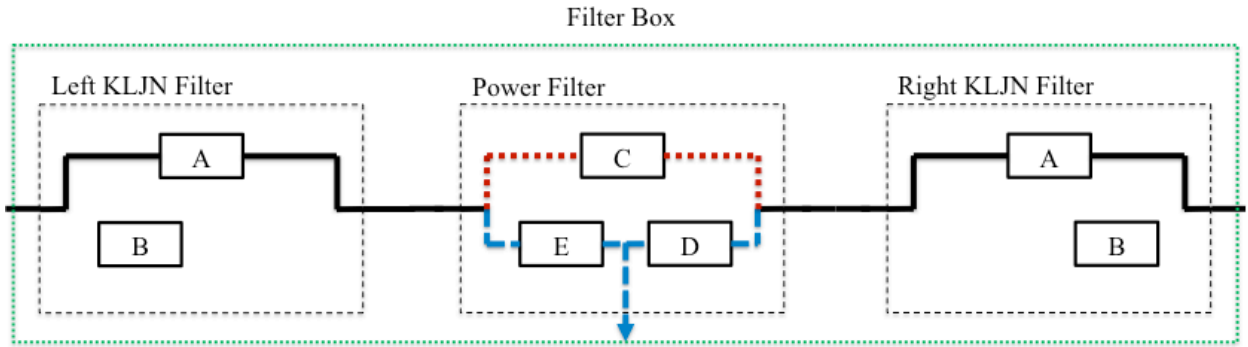

**Figure 5.** The filter box of the inactive host (when it is not executing KLJN key exchange): State 1. Everything is passing between left and right and the host can access only the power. Filter A is passing everything (shorted). Filter B is disconnected. Filter C is passing $B_{kljn}$ only. Filters and E and D are passing $f_p$ only. State 1 is when the host is not allowed to access KLJN band. State 2 is when the host is allowed to access KLJN band. This filter box is in State 1.

| KLJN Filters | Filter A | Filter B |
|---|---|---|
| KLJN $B_{kljn}$ Allowed | Yes | No |
| Power Frequency Allowed | Yes | No |

Table 1. Truth table of the KLJN Filters in State 1 (inactive host).

| Power Filter | Filter C | Filter D | Filter E |
|---|---|---|---|
| KLJN $B_{kljn}$ Allowed | Yes | No | No |
| Power Frequency Allowed | No | Yes | Yes |

Table 2. Truth table of the Power Filter in State 1 (inactive host).

In Figure 7 there are seven key exchanges occurring simultaneously with every host in that network active and allowed access to the KLJN band. Moreover, the power filters of these hosts must separate the KLJN loops by rejecting $B_{kljn}$. We call this working mode of the filter boxes of hosts executing key exchange *State 2*. The wiring and frequency transfer of the Filter Box in State 2 are shown in Figure 6 and Tables 3,4.

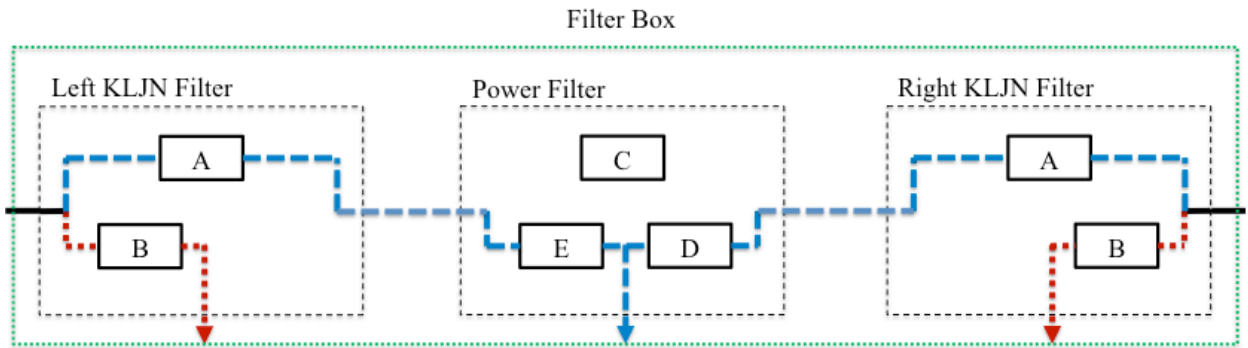

**Figure 6.** The filter box of the active host (when it is executing a KLJN key exchange): State 2. The power is passing between left and right but the KLJN band is not and the left and right KLJN units are separated while doing a key exchange toward left and right. State 1 is when the host is not allowed to access KLJN band. State 2 is when the host is allowed to access KLJN band. This filter box is in State 2.

| KLJN Filter | Filter A | Filter B |
|---|---|---|
| $B_{kljn}$ allowed | No | Yes |
| $f_p$ allowed | Yes | No |

Table 3. Truth table of left KLJN filter when a host is in State 2 (active host).

| Power Filter | Filter C | Filter D | Filter E |
|---|---|---|---|
| $B_{kljn}$ allowed | No | No | No |
| $f_p$ allowed | No | Yes | Yes |

Table 4. Truth table of power filter when a host is in State 2 (active host).

In this section we have shown that the line can be packed with non-overlapping KLJN loops to execute simultaneous key exchanges between selected hosts. In the next section, we propose a network protocol to provide secure keys for each host to be able to communicate securely via the internet or other publicly accessible channels between arbitrary pairs of hosts. The time requirement of key exchange over the whole smart grid will also be analyzed versus the network size $N$.

## 2.2. Protocol and Speed

To quickly and efficiently connect every host with all other host in the same one-dimensional network we need to establish a protocol. The protocol must make every possible connection in the network, must not overlap loops, and must be quick and efficient by making as many simultaneous loops without overlapping as possible.

To determine the time and speed requirements to establish a KLJN secure key exchange we must first define terms. In the classical KLJN system, where only the noise existed in the wire, the low-frequency cutoff of the noise was 0 Hz and the high-frequency cut-off was $B_{kljn}$. In the case of KLJN in a smart grid, this situation will be different because of the power frequency. However, at short distances (less than 10 miles), the $B_{kljn}$ band can be beyond the power frequency $f_p$ and the difference is negligible. Then the shortest characteristic time in the system is the correlation time $\tau_{kljn}$ of the noise ($\tau_{kljn} \approx 1/B_{kljn}$). $B_{kljn}$ is determined by the distance $L$ between Alice and Bob so that $B_{kljn} << c/L$ [21] (for example, $B_{kljn} << 100\,\text{kHz}$ for $L$=1 kilometer). Alice and Bob must make a statistics on the noise, which typically requires around 100 $\tau_{kljn}$ duration [25] (or 0.01 seconds if we use $B_{kljn} = 10\,\text{kHz}$) to have a sufficiently high fidelity (note, faster performance is expected in advanced KLJN methods [31]). A bit exchange (BE) occurs when Alice and Bob have different resistor values, this occurs on average of 200 $\tau_{kljn}$ or 0.02 seconds if $B_{kljn} = 10\text{kHz}$. The length of the secure key exchange can be any arbitrary length. For example if we have a key length of 100 bits then, we need 100 BE which requires on average 20000 $\tau_{kljn}$ which is approximately 2 seconds if $B_{kljn}$ is 10 kHz. Once the KLJN secure key has been exchanged the total amount of time needed to complete this is one KLJN secure key Exchange period (KE). While the key exchange is slow, the system has the advantage that it is running continuously (not only during the handshake period like during common secure internet protocols) thus large number of secure key bits are produced during the continuous operation.

For the sake of simplicity, we use the pessimistic estimation by assuming a uniform duration for KE determined by the largest distance in the network even though short distances can exchange keys at a higher speed.

The protocol we propose here first connects the nearest neighbor of every host; this allows the highest number of simultaneous non-overlapping loops per KE and only requires one KE to complete this first step. The protocol then connects the second nearest neighbors; this allows the second highest numbers of simultaneous loops per KE. However, due to the requirement of avoiding overlapping loops, connecting each pairs of second nearest neighbors requires two KEs. The protocol then connects the third nearest neighbors which requires 3 KEs to complete and connects the third most simultaneous loops per KE. This procedure continues until the *i*-th nearest neighbor is equal to or less than half of the size of the network. If the number of steps $i$ between the *i*-th nearest neighbors satisfies the relation $i > N/2$ then, to avoid overlapping loops, only one connection per KE is possible.

As an example, we will show in the next section that for $N = 7$ (see Figure 2) 16 KEs (or approximately 32 seconds if $B_{kljn}$ is 10 kHz) are required when the keys are 100 bits long. Using this protocol, the analytic form of the exact time required to fully arm every host with enough keys to securely communicate with everybody in the network is dependent on the size of the network and whether the network has an even or odd size. In the following sections we will deduce the analytic relations and show examples.

### 2.2.1 The network size *N* is an odd number

We illustrate the calculation of time requirement with examples shown in the following figures. A general formula for an arbitrary size network when *N* is odd is given later. In this example we have a network of size $N = 7$. We have 8 host with index $i$ $(0 \leq i \leq 7)$. We have 7 intermediate connections between the first and last host.

The first step in the protocol connects the nearest neighbors, see Figure 7.

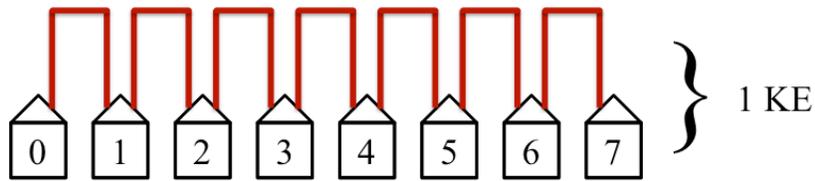

**Figure 7.** The first step in the protocol connects the nearest neighbors. This step is the quickest and most efficient. It has the most non-overlapping simultaneous loops and requires only 1 KE to complete. Every host in this step has access to KLJN band and thus are in State 2.

The second step in the protocol will then connect the second nearest neighbors, see Figure 8.

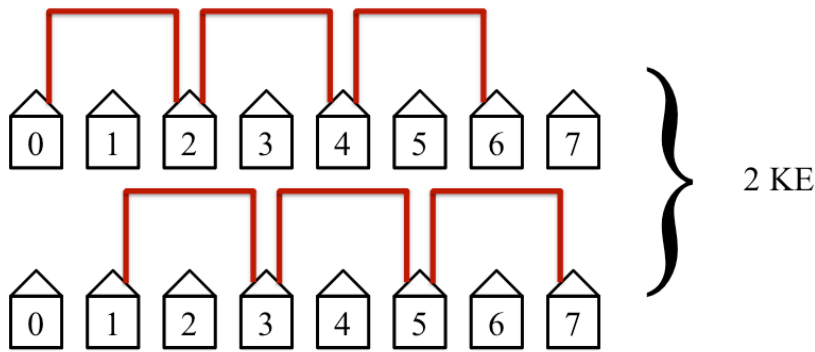

**Figure 8.** The second step in the protocol connects the second nearest neighbors. This step is the second quickest and the second most efficient. It has the second most non-overlapping simultaneous loops and requires 2 KEs to complete.

The protocol will then connect the third closest neighbors as shown in Figure 9. This will take 3 KEs to complete and is not as efficient as the first two steps in the protocol but still has simultaneous loops in two of its KE steps.

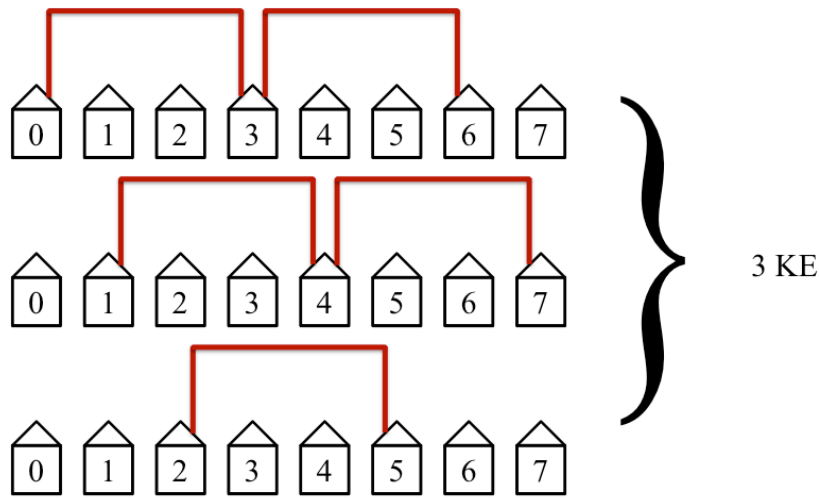

**Figure 9.** The third step in the protocol connects the third nearest neighbors. This step is not as efficient as the first two steps but still has simultaneous loops. This step requires 3 KEs to complete.

The protocol will then connect the fourth nearest neighbors as shown in Figure 10. This is above the midpoint for our example with $N = 7$ and is the slowest and least efficient step in the protocol. The midpoint is considered when the distance between Alice and Bob is equal to half the length of the network. These steps will take 4 KEs to complete. Simultaneous loops with disconnected hosts are no longer possible beyond the midpoint. The slowest and least efficient steps occur at the midpoint of the protocol.

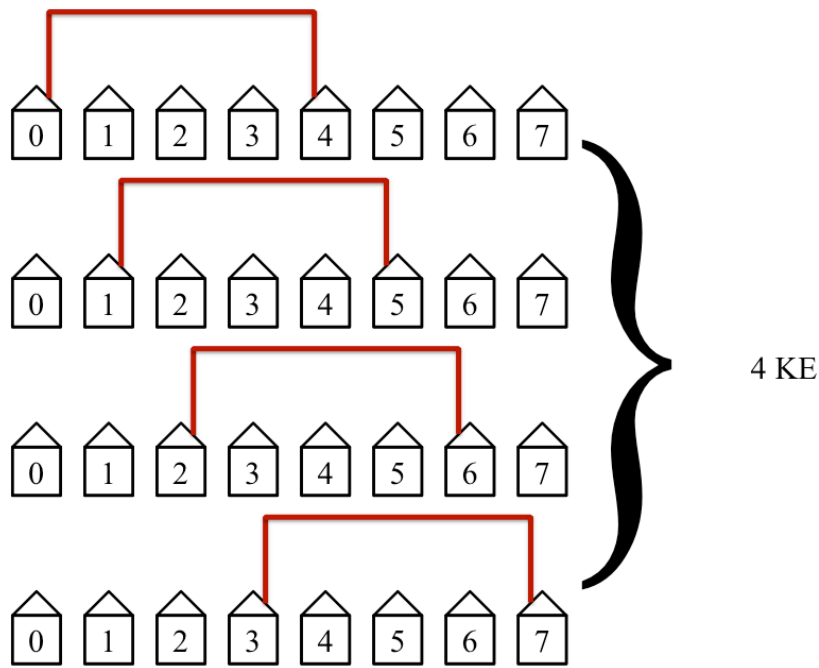

**Figure 10.** The fourth step in the protocol connects the fourth nearest neighbors. This step is the slowest and least efficient step in the protocol in our example of $N = 7$. This step requires 4 KEs to complete.

The protocol will then connect the fifth nearest neighbors as shown in Figure 11. This step will take 3 KEs to complete. It is also inefficient since it is beyond the midpoint thus only a single loop is possible, but it requires fewer KEs since there are only three such pairs.

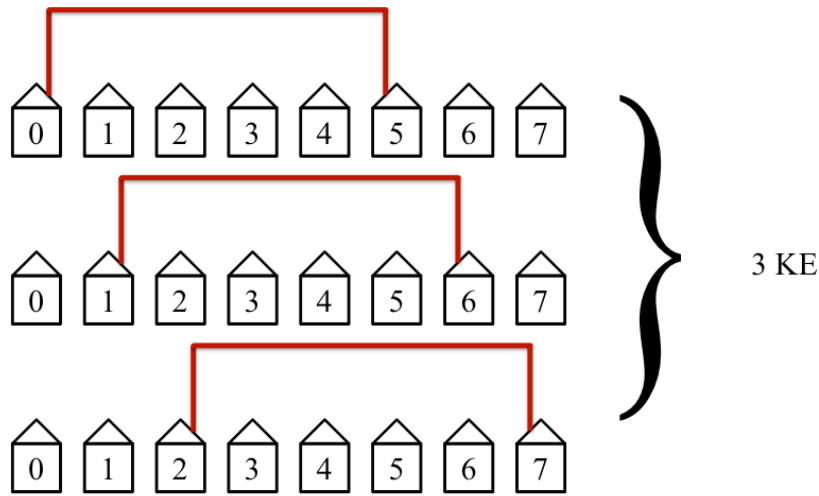

**Figure 11.** The fifth step in the protocol connects the fifth nearest neighbors. This step is not efficient since simultaneous non-overlapping loops with disconnected hosts cannot occur.

The protocol will then connect the sixth nearest neighbors as shown in Figure 12. This step will take 2 KEs because there are only two possibilities.

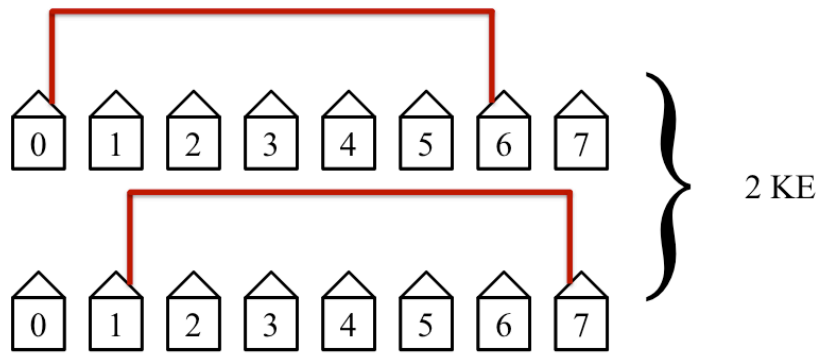

**Figure 12.** The sixth step in the protocol connects the sixth nearest neighbors. This step requires only 2 KEs since there are only two possibilities.

The protocol will then connect the seventh closest neighbors as shown in Figure 13. This will take 1 KE since there is only one such pair of hosts.

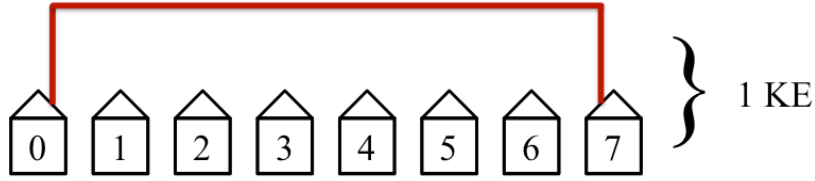

**Figure 13.** Only one key exchange is performed in this step. Host 1 through 6 are not allowed access to KLJN band thus they are in State 1. The seventh and the last step. This step is not efficient but only requires one KE since there is only one such pair of hosts.

This completes the protocol for an example of size $N = 7$. Notice the pattern that occurs for $N$ being odd. We have a pattern of 1 KE, 2 KE, 3 KE, 4 KE, 3 KE, 2 KE, and 1 KE. This is essentially Gauss's counting technique up to $N/2$ and back. The total number of KEs needed will be $1\text{KE} + 2\text{KE} + 3\text{KE} + 4\text{KE} + 3\text{KE} + 2\text{KE} + 1\text{KE} = 16\text{KE}$.

The speed or time requirement of the protocol for a network of arbitrary size $N$ with $N$ being odd is $\left(\dfrac{N+1}{2}\right)^2$ KEs and can be derived as follows.

Since $N$ is odd we can express it as;

$$N = 2n+1. \tag{2}$$

To find the midpoint we can solve $n$ and express it in terms of $N$, this gives the following;

$$\frac{N-1}{2} = n. \tag{3}$$

The pattern when $N$ is odd has the following form;

$$1 + 2 + \cdots + (n-1) + n + (n-1) + \cdots + 2 + 1 = \left(\frac{N-1}{2}\right)^2. \tag{4}$$

Expressing $n$ in terms of $N$ gives;

$$1 + 2 + \cdots + \left(\frac{N-1}{2} - 1\right) + \left(\frac{N-1}{2}\right) + \left(\frac{N-1}{2} - 1\right) + \cdots + 2 + 1 = \left(\frac{N-1}{2}\right)^2. \tag{5}$$

We know from Gauss's counting method that,

$$1 + 2 + \cdots + N = \frac{N(N+1)}{2}. \tag{6}$$

In our pattern we can use Gauss's counting method twice to find the sum as follows.

$$1+2+\cdots+\underbrace{\left(\frac{N-1}{2}-1\right)}_{\frac{\left(\frac{N-1}{2}-1\right)\left(\frac{N-1}{2}\right)}{2}}+\left(\frac{N-1}{2}\right)+\underbrace{\left(\frac{N-1}{2}-1\right)+\cdots+2+1}_{\frac{\left(\frac{N-1}{2}-1\right)\left(\frac{N-1}{2}\right)}{2}}=\left(\frac{N-1}{2}\right)^2.$$

(7)

This simplifies to

$$\left(\frac{\left(\frac{N-1}{2}\right)\left(\frac{N-1}{2}-1\right)}{2}\right)+\left(\frac{N-1}{2}\right)+\left(\frac{\left(\frac{N-1}{2}\right)\left(\frac{N-1}{2}-1\right)}{2}\right)=\left(\frac{N-1}{2}\right)^2.$$

(8)

Thus the speed of the network is proportional to $\frac{N^2}{4}$ with $N$ being odd and the size of the network. The pattern for when $N$ is even is similar.

### 2.2.2 The network size $N$ is an even number

For the sake of easier understanding and for those without a communications background, we will again illustrate the calculation of time requirement with examples shown in the following figures. In this case, we have an even number as network size $N = 8$. We have 9 host with index $i$ $(0 \leq i \leq 8)$. We have 8 intermediate connections between the first and last host.

The first step in the protocol connects the nearest neighbors. This step is the quickest and most efficient. It has the most simultaneous non-overlapping loops and requires only one KE to complete. Figure 14 illustrates this first step in the protocol.

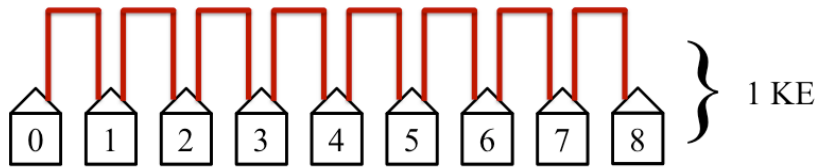

**Figure 14.** The first step in the protocol connects the nearest neighbors. This step is the quickest and most efficient. It has the most non-overlapping simultaneous loops and requires only 1 KE to complete.

The second step in the protocol will then connect the second nearest neighbors as shown in Figure 15. This step will take two KEs to complete and has the second most simultaneous non-lapping loops. It is the second quickest and second most efficient step.

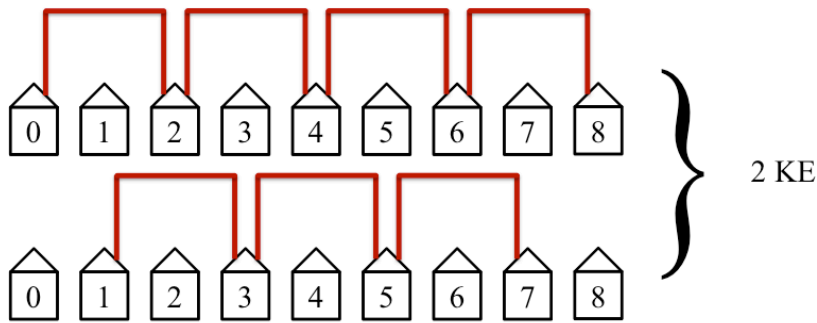

**Figure 15.** The second step in the protocol connects the second nearest neighbors. This step requires 2 KEs to complete.

The protocol will then connect the third nearest neighbors as shown in Figure 16. This will take 3 KEs to complete and is not as efficient as the first two steps in the protocol but still has simultaneous loops in this example of $N = 8$.

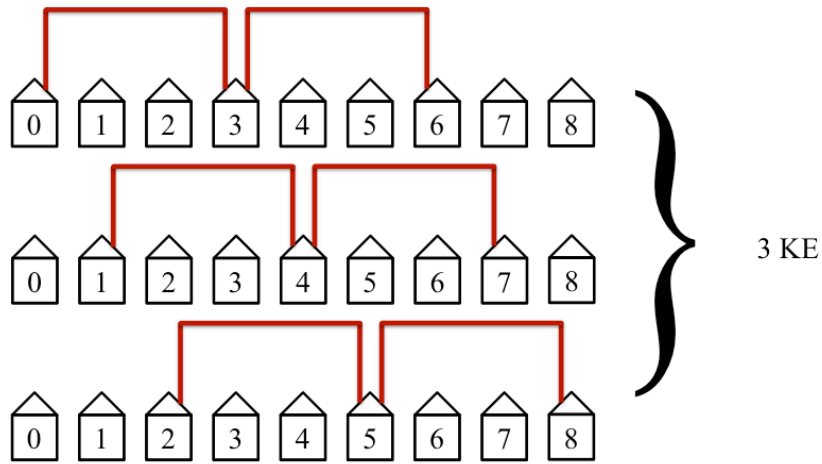

**Figure 16.** The third step in the protocol connects the third nearest neighbors. This step requires 3 KEs to complete.

The protocol will then connect the fourth nearest neighbors as shown in Figure 17. This is at the midpoint for our example with $N = 8$ and is the slowest and least efficient step in the protocol. The midpoint is defined when the distance between Alice and Bob is equal to half the length of the network. This step will take 4 KEs to complete. The slowest and least efficient steps occur at the midpoint of the protocol.

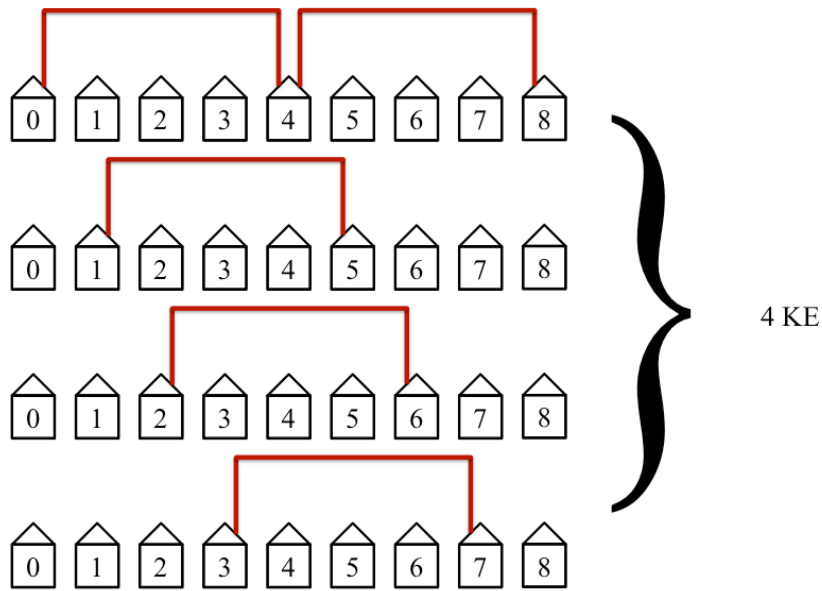

**Figure 17.** The fourth step in the protocol connects the fourth nearest neighbors. It requires 4 KEs to complete.

The protocol will then connect the fifth nearest neighbors as shown in Figure 18. This step will take 4 KEs to complete. It is not efficient since it is at midpoint.

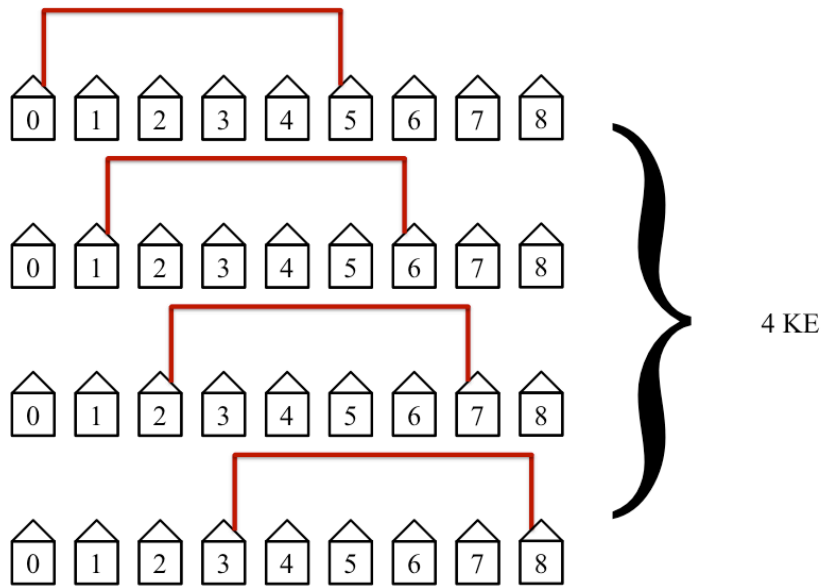

**Figure 18.** The fifth step in the protocol connects the fifth nearest neighbors. This step is not efficient since simultaneous non-overlapping loops with disconnected hosts cannot occur. It requires 4 KEs to complete.

The protocol will then connect the sixth nearest neighbors as shown in Figure 19. This step will take 3 KEs because there are only three possibilities at this distance in this example of a network of size $N = 8$.

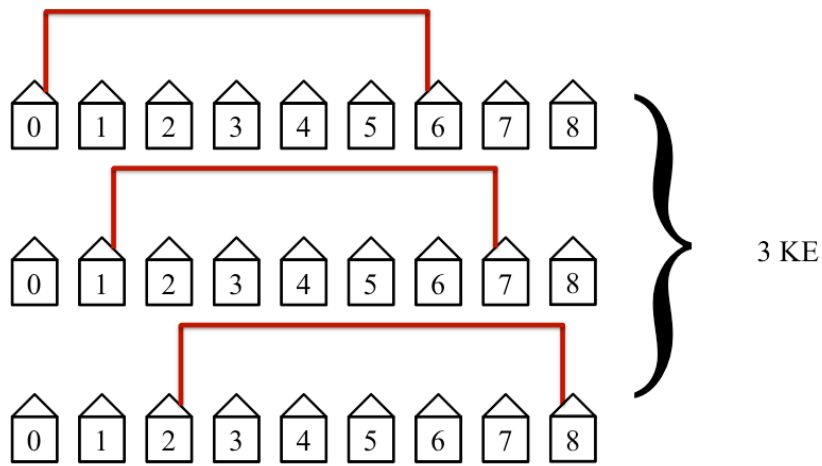

**Figure 19.** The sixth step in the protocol connects the sixth nearest neighbors. This step requires only 3 KEs since it is the third to last step and there are only three possibilities.

The protocol will then connect the seventh nearest neighbors as shown in Figure 20. This will take 2 KEs since there are only two pairs of host with a length of seven hosts between them.

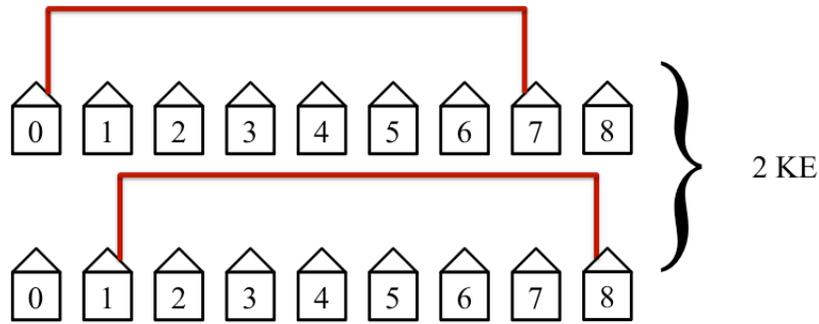

**Figure 20.** The seventh step in this example of a network of size $N=8$. This step is not efficient but only requires two KEs since there are only two such pairs of host.

The last step in the protocol connects the first and last hosts. This step is the least efficient and requires the entire length of the network. Since there is only one pair of host at this length, this step requires only one KE. This last step is illustrated in Figure 21.

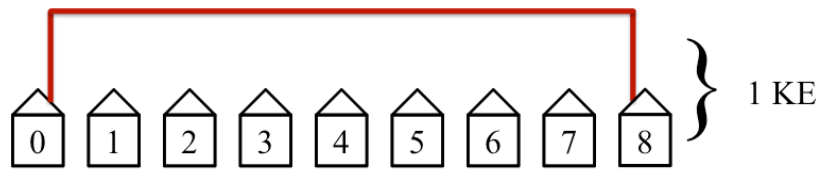

**Figure 21.** The last step in our example with $N=8$. This step is not efficient but only requires one KE since there is only one pair of hosts that are eight hosts apart.

Notice the pattern that occurs for $N$ being even. We have 1 KE, 2 KE, 3 KE, 4 KE, 4 KE, 3 KE, 2 KE, and 1 KE. This is essentially Gauss's counting technique up to $N/2$ and back. The total number of KEs needed will be $1\text{KE}+2\text{KE}+3\text{KE}+4\text{KE}+4\text{KE}+3\text{KE}+2\text{KE}+1\text{KE}=20\text{KE}$. The time needed to connect the entire network will take 20 KEs which is approximately 40 seconds if $B_{kljn}$ is 10 kHz and if the key is 100 bits long.

The speed or time requirement of the protocol for a network of size $N$ with $N$ being even between the first and last host is $\dfrac{N^2}{4}+\dfrac{N}{2}$ KEs and can be derived as follows.

With $N=8$ the pattern in our example is;

$$\frac{N^2}{4}+\frac{N}{2}=20\text{KE} .\qquad(9)$$

Since $N$ is even we can express it as;

$$N=2n .\qquad(10)$$

To find the midpoint we can solve *n* and express it in terms of *N*, this gives the following;

$$\frac{N}{2} = n. \tag{11}$$

The general pattern when *N* is even has the following form;

$$1 + 2 + \cdots + n + n + \cdots + 2 + 1 = \frac{N^2}{4} + \frac{N}{2}. \tag{12}$$

Expressing *n* in terms of *N* gives;

$$1 + 2 + \cdots + \frac{N}{2} + \frac{N}{2} + \cdots + 2 + 1 = \frac{N^2}{4} + \frac{N}{2}. \tag{13}$$

We know from Gauss's counting method that,

$$1 + 2 + \cdots + N = \frac{N(N+1)}{2}. \tag{14}$$

In our pattern we can use Gauss's counting method twice to find the sum as follows.

$$\underbrace{1 + 2 + \cdots + \frac{N}{2}}_{\frac{\left(\frac{N}{2}\right)\left(\frac{N}{2}+1\right)}{2}} + \underbrace{\frac{N}{2} + \cdots + 2 + 1}_{\frac{\left(\frac{N}{2}\right)\left(\frac{N}{2}+1\right)}{2}} = \frac{N^2}{4} + \frac{N}{2}. \tag{15}$$

$$\frac{\frac{N}{2}\left(\frac{N}{2}+1\right)}{2} + \frac{\frac{N}{2}\left(\frac{N}{2}+1\right)}{2} = \frac{N^2}{4} + \frac{N}{2}. \tag{16}$$

This simplifies to

$$\left(\frac{N}{2}\right)\left(\frac{N}{2}+1\right) = \frac{N^2}{4} + \frac{N}{2}. \tag{17}$$

Thus the speed of the network is proportional to $\frac{N^2}{4}$ with *N* being the size of the network and even.

# Limitations of the study, open questions, and future work

To fully implement the KLJN key exchange protocol over the grid will require solutions to further engineering problems. This paper presents results of our early work, which focuses on the system-concept in a one-dimensional network. Some of the limitations, open questions, and future work are discussed below.

## 3.1. Limitations

The main limitation of the KLJN protocol is that it is a peer-to-peer network. This will limit the number of simultaneous KLJN key exchanges a host can have. Since overlapping loops are not allowed, the time required scales quadratically with the number of hosts. Another limit is that the KLJN bandwidth is dependent on the distance between Alice and Bob and slows down for large distances. These limitations make it impractical to connect millions of hosts via a linear chain thus other topologies (and perhaps bridges or routers) will be needed to connect such chains with each other. Practical limitations in the power system, such as tap-changing transformers and other devices may also require bridges to couple the KLJN signal around these devices.

## 3.2. Open questions

The related technical challenges need further researches that are straightforward. For example distribution transformers can shield most of the signals sent from one phase from the load side but there are many ways to get around them with the KLJN band. We don't investigate the problems of phase-correcting inductors and capacitors since they are separated by the power filters from the KLJN band. Research and development will be needed for some of these problems including how to setup filters in each node. Accuracies are typically within a few percent. In the experimental demo the cable resistance was 2% of total loop resistance. In practice, the impedance of the power grid would need to be taken into account.

## 3.3. Future work

Future work will, among others, include protocols for several other power grid topologies. Setting up filters on the power grid and implementing all the filters will also be studied. Hacking attacks against filters and the defense against them is also an interesting open problem.

# Conclusions

We have introduced a protocol to offer unconditionally secure key exchange over the smart grid. We used a reconfigurable filter system and proposed a special protocol to create non-overlapping single loops in the smart power grid for the realization of the Kirchhoff-Law-Johnson-(like)-Noise secure key distribution system. The protocol is valid for one-dimensional networks like the radial electrical power distributions grid. We carried out a scaling analysis for the speed of the protocol versus the size of the grid. When properly generalized, such a system has the potential to achieve unconditionally secure key distribution over the smart power grid of arbitrary dimensions.

Before the implementation of the protocol can take place several practical questions must be answered such as the impact of finite and possibly varying wire resistance, capacitance and power load on the security; and the applications of relevant privacy amplification methods. Other questions include changing size $N$, hacking attacks against the filter control and the relevant defense tools. We also discussed the limitations of the KLJN key exchange protocol, open questions surrounding the implementation on the smart grid, and future work required. Since this paper is a system-concept study we leave the details to future work.

# Acknowledgements

Discussions with Mladen Kezunovic and Karen Butler-Perry about smart grid security are appreciated. LK is grateful to Horace Yuen for discussions about physical secure key exchange systems. Discussions with Yessica Saez are also appreciated.